\documentstyle[12pt,amstex,amssymb,amscd]{article}
\def\tS{\tilde{S}}
\def\to{\rightarrow}
\def\np#1{{\sl Nucl.~Phys.~\bf B#1}}

\def\pr#1{{\sl Phys.~Rev.~\bf D#1}}

\def\Kmax{K_{\rm max}}\def\rQCD{{\rm QCD}}
\let\sstl=\scriptscriptstyle

\reversemarginpar
\allowdisplaybreaks
\begin{document}
 
\begin{titlepage}
 
\begin{flushright}
{\bf 
     UTHEP-95-1101\\
     Nov.,~1995\hspace*{0.47in}}
\end{flushright}

\vspace{0.1cm}
\begin{center}
{\LARGE
Dokshitzer-Gribov-Lipatov-Altarelli-Parisi Evolution and 
the Renormalization Group
Improved Yennie-Frautschi-Suura Theory in QCD$^{\S}$
}
\end{center}
 
\begin{center}
 {\bf B.F.L. Ward}\\
   {\em Department of Physics and Astronomy,\\
   The University of Tennessee, Knoxville, Tennessee 37996-1200\\
   SLAC, Stanford University, Stanford, California 94309}\\
   and \\
   {\em CERN, Theory Division, CH-1211 Geneva 23, Switzerland}\\
{\bf S. Jadach}\\
   {\em Institute of Nuclear Physics,
        ul. Kawiory 26a, Krak\'ow, Poland}\\
   {\em CERN, Theory Division, CH-1211 Geneva 23, Switzerland,}\\
\end{center}

\vspace{1.0cm}
\abstract{
We show that the recently derived 
renormalization group improved Yennie-Frautschi-Suura (YFS)
exponentiation of soft $(k_o\rightarrow 0)$ gluons in QCD is fully
compatible with the usual 
Dokshitzer-Gribov-Lipatov-Altarelli-Parisi (DGLAP) evolution
of the structure functions of hadrons in the respective 
hadron-hadron hard interactions. We show how to implement
the YFS exponentiation without double or over counting effects
already implied by the DGLAP equation. In this way, we arrive at a theory
which allows for the development
of realistic, multiple gluon Monte Carlo event generators
for hard hadron-hadron scattering processes in which the DGLAP
evolved structure functions are correctly synthesized with the
respective YFS exponentiated soft gluon effects in a rigorous way.
\par
}
 
\begin{center}
{\it Submitted to Phys. Lett. B}
\end{center}
 
\vspace{0.5cm}
\renewcommand{\baselinestretch}{0.1}
\footnoterule
\noindent
{\footnotesize
\begin{itemize}
\item[${\S}$]
Work supported in part by the US DoE contract DE-FG05-91ER40627,
Polish Government grants KBN 2P30225206 and 2P03B17210
and Polish-French Collaboration
within IN2P3.
\end{itemize}
}
 
\begin{flushleft}
{\bf 
     UTHEP-95-1101\\
     November  1995 }
\end{flushleft}
 
\end{titlepage}
 
 
An important step in the implementation of the recently 
derived~\cite{qcdexp} renormalization group improved
soft Yennie-Frautschi-Suura (YFS)\cite{yfs}
gluon exponentiation in QCD is the proper synthesization of these new
infrared (IR) effects ( where by infrared we do mean the gluon energy $k_o
\rightarrow 0$ limit as opposed to the soft gluon regime analyzed by
Dokshitzer {\it et al.}\cite{ddmt} in which $k_{\perp}\ge {1\over R}$,
$R\simeq$ hadron radius $\simeq 1/(200MeV)\simeq {1\over \Lambda_{QCD}}$) with
the required Dokshitzer-Gribov-Lipatov-Altarelli-Parisi (DGLAP)~\cite{DGLAP}
evolution of the
respective hadron structure functions in computing any particular
hard hadron-hadron scattering cross section, such as 
$p+\bar p\rightarrow t\bar t+X$, for example. Indeed, a naive use of the 
results in Ref.~\cite{qcdexp} would lead to double or over counting
some of the same effects, rendering the precision of any such calculated
cross section questionable at best. In what follows, we will show how
one properly synthesizes the DGLAP structure function evolution with
renormalization group improved soft gluon YFS QCD exponentiation
derived in Ref.~\cite{qcdexp}. In this way, we develope the necessary
remaining theoretical apparatus needed for the construction
of realistic hadron-hadron hard scattering Monte Carlo event generators
in which finite $P_{\perp}$ multiple soft gluon effects are taken
into account at the level of the hard scattering subprocess amplitude
while the proper structure function evolution is realized, on an 
event-by-event basis. Such a Monte Carlo event generator will appear
elsewhere~\cite{elswh}.\par
Specifically, let us recall the basic result from Ref.~\cite{qcdexp}
for the YFS exponentiated differential cross section for 
the process 
$q+{\llap{\phantom q}^{\sstl(}\bar q^{\sstl)}{}}'\to q''+{\llap{\phantom q}^{\sstl(}\bar q^{\sstl)}{}}'''+n(G)$, where G is a soft gluon,
\begin{eqnarray}d\hat\sigma_{exp}&=exp[{\rm SUM_{IR}(QCD)}]\sum_{n=0}^\infty{
1\over n!}\int\prod_{j=1}^n{d^3k_j\over k_j}\int{d^4y\over(2\pi)^4}e^{+iy(p_1+
p_2-q_1-q_2-\sum_jk_j)+D_{QCD}} \nonumber \\
&\times \bar\beta_n(k_1,\dots,k_n){d^3q_1d^3q_2\over q_1^{\,0}
q_2^{\,0}}\label{eone}\end{eqnarray} where\begin{equation}
D_{QCD}=\int{d^3k\over
k_0}\tilde S_{\rm QCD}\left[e^{-iy\cdot k}-\theta(K_{max}-|\vec k|)\right]\quad.
\label{etwo}
\end{equation}
Here, we have defined 
\begin{equation}{
\rm SUM_{IR}(QCD)}\equiv2\alpha_s\tilde B_{\rm QCD}+2\alpha_s\Re B_{\rm QCD}
\label{ethree}\end{equation}
from Ref.~\cite{qcdexp} with the QCD YFS virtual and real genuinely
infrared functions $\tilde B_{\rm QCD}$, $B_{\rm QCD}$, and 
$\tilde S_{\rm QCD}$
as defined therein\cite{qcdexp}. The hard gluon residuals $\bar\beta_n(k_1,\dots,k_n)$ are also as defined in Ref.~\cite{qcdexp}, for example.
At issue is how one synthesizes a calculation such as that in 
(\ref{eone}) with the fundamental structure functions for the incoming
partons which are themselves solutions 
of the DGLAP equation and hence in general
may have in themselves some of the effects in (\ref{eone}) 
already included?\par
Indeed, one might think that the usual formula
\begin{equation}
\sigma(p\bar p\rightarrow t\bar t+X) = 
\int\sum_{i,j}D_i(x_i)\bar D_j(x_j)d\hat\sigma_{exp}(x_ix_js)dx_idx_j
\label{efour}
\end{equation}
could be used to leading log accuracy when the existing structure functions
$D_i(x_i),\bar D_j(x_j)$ for partons $i,j$ in $p,\bar p$, respectively,
in the literature\cite{strfn} are used. Here, $x_i,x_j$ are the
respective light-cone momentum fractions.
But, this is not even true
because the DGLAP equation is well known to contain all leading logs
from the initial state radiation of gluons and all attendant collinear
infrared effects. Thus, in using (\ref{efour}) naively one would
double count some leading log and collinear infrared effects.\par
Specifically, in order to use the existing experience on the 
structure functions together with the YFS exponentiation improvement
of perturbation theory given by (\ref{eone}), one has to identify
those effects in the YFS formula which are already contained in 
the DGLAP equation for the structure functions and remove them. 
This we now do.\par
Since we have exponentiated the quantity $SUM_{IR}(QCD)$ in (\ref{eone}),
we start by isolating from it and from all of its attendant parts in
(\ref{eone}) that collinear infrared contribution which is generated by the 
DGLAP evolution of the structure functions. Viewing the DGLAP evolution
as the complete LL series from the initial state radiative effects,
we see that we must do the following to synthesize (\ref{eone})
with the existing structure functions~\cite{strfn} in (\ref{efour}):
\begin{itemize}
\item From the initial state part of $\tS_{QCD}$, we must
remove that part which corresponds to the YFS limit of the DGLAP kernel,
$\tS_{QCD}|_{DGLAP}$. This can be computed using the definitions 
in the original paper of Altarelli and Parisi\cite{DGLAP} and we get
the result that, if we wish to use (\ref{eone}) in (\ref{efour}),
we should replace $\tS_{QCD}$ in (\ref{eone}) with
${\tS_{QCD}}^{nls}$ where
\begin{equation}
{\tS_{QCD}}^{nls} = \tS_{QCD} -\tS_{QCD}|_{DGLAP}
\label{efive}
\end{equation}
for 
\begin{align}
\tS_{QCD}|_{DGLAP}(k) &= {C_F\alpha_s\over 4\pi^2}{\Big [} {\theta(z)~4(1-zv)^2\over (k_\bot^2+z^2v^2m_q^2)}{\Big\lgroup} 1-{m_q^2v^2z^2\over k_\bot^2+z^2v^2m_q^2}{\Big\rgroup} \cr
                      &~~+{\theta(-z)~4(1+zv)^2\over (k_\bot^2+z^2v^2m_q^2)}{\Big\lgroup} 1-{m_q^2v^2z^2\over k_\bot^2+z^2v^2m_q^2}{\Big\rgroup} {\Big ]}
\label{esix}
\end{align}
with the identifications $zk=k_z$, $k^o=v\sqrt s/2$,$\vec k=(k_z,\vec k_\bot)$
and we have written the RHS of (\ref{esix}) for incoming $q$ and $\bar q$ 
respectively for definiteness; for the $q\bar q'$ incoming case, for example,
one would need to substitute $m_{q'}$ for $m_q$ in the coefficient of 
$\theta(-z)$ on the RHS of (\ref{esix}). The superscript $nls$ on $\tS_{QCD}$
on the LHS of (\ref{efive}) denotes that it is non-leading in the s-channel
relative to DGLAP evolution. Thus,  $\tS_{QCD}^{nls}$ has the property that
its integral over the gluon phase space has no collinear big logarithm
terms proportional to ${\alpha_s L\over \pi}$ or ${\alpha_s L^2 \over \pi}$
, where $L= \ln s/m_q^2-1$ in the case of incoming $q\bar q$, for example.
To obtain the result (\ref{esix}), one follows the derivation of the
DGLAP kernel $P_{Gq}(\bar z)$ in the paper of Altarelli and Parisi in
Ref.~\cite{DGLAP} but one substitutes the YFS vertex  $-ig\tau^a(2p+k)_\mu$
instead of the complete vertex $-ig\tau^a\gamma_\mu$ and one retains the
quark mass corrections through order $m_q^2$ in the respective kinematics.
Here, $g$ is the QCD coupling constant and $\tau^a$ generate the quark
(vector) representation of the QCD gauge group; $p$ is the final
4-momentum of the respective radiating quark.
We stress that ${\tS_{QCD}}^{nls}$ defined in this way is positive
definite.
The result (\ref{efive}), which corresponds to the replacement
$\tilde B_{QCD}(K_{max})\rightarrow \tilde B_{QCD}^{nls}(K_{max})=
\int {d^3k\over k_o}\theta(K_{max}-|\vec k|)\tS_{QCD}^{nls}$,
then avoids the double counting of real gluon radiation from the
initial state that is already included in the DGLAP evolution of the
structure functions.\par
\item
This brings us to the possible double counting of the virtual corrections
represented by $\Re B_{QCD}$ in (\ref{eone}). We address this problem as
follows. The DGLAP evolution equation generates the entire leading log
series associated with initial state real and virtual gluon radiation.
Thus, we should also remove from $\Re B_{QCD}$ all leading log effects
associated with the initial state. Here, we stress that the definition
of the big log $L$ has to be the same as that used in the DGLAP evolution.
In practice, this means that, in the
formula for $SUM_{IR}(QCD)$ in eq.(9) of Ref.~\cite{qcdexp}, for the
initial state contribution corresponding to the case $A=s$ in the 
respective sum over channels in this formula, we should use
the result 
\begin{equation}
B_{tot}(s)= {\pi^2\over 3} -{1\over 2}
\label{eseven}
\end{equation}
instead of the result given by eq.(10) in Ref.~\cite{qcdexp}.
We denote this new result by  $SUM^{nls}_{IR}(QCD)$. 
In addition, in the computation of the hard gluon residuals $\bar\beta_n$
in (\ref{eone}) one should remember again to remove from the real part of the
integral over the first term in eq.(4) in Ref.~\cite{qcdexp} (this
integral can be found in Refs.~\cite{yfsint}), which is the initial state
part of $\Re B_{\rm QCD}$,
all terms involving
either ${\alpha_s L\over \pi}$ and ${\alpha_s L^2\over \pi}$, where
$L\equiv \ln s/m_q^2 -1$ here in the $q\bar q$ incoming case, for example.
The generalization to the $q\bar q'$ incoming case is immediate.
Henceforth, we denote this new $\Re B_{QCD}$ with its initial state
s-channel big logarithms removed as $\Re B^{nls}_{QCD}$.
It is $\tS^{nls}_{QCD}$ and $\Re B^{nls}_{QCD}$ 
that should now be used in computing 
$\bar\beta_n$ in (\ref{eone}); we denote the resulting new hard gluon residual
as $\bar\beta^{nls}_n$,  where it is understood
that the various finite order perturbative QCD differential cross sections
in this definition of $\bar\beta^{nls}_n$
are the reduced, hard mass factorized cross sections.\par
\end{itemize}
When these steps are followed, we arrive at a new
representation of the cross section in (\ref{eone}) in
which $\tS^{nls}_{QCD}$, $SUM^{nls}_{IR}(QCD)$ and $\Re B^{nls}_{QCD}$
are substituted for $\tS_{QCD}$, $SUM_{IR}(QCD)$ and $\Re B_{QCD}$
respectively 
and
we write this new version of the result in (\ref{eone}) as
\begin{eqnarray}d\hat\sigma'_{exp}&=exp[{\rm SUM^{nls}_{IR}(QCD)}]\sum_{n=0}^\infty{1\over n!}\int\prod_{j=1}^n{d^3k_j\over k_j}\int{d^4y\over(2\pi)^4}e^{+iy(p_1+
p_2-q_1-q_2-\sum_jk_j)+D^{nls}_{QCD}} \nonumber \\
&\times \bar\beta^{nls}_n(k_1,\dots,k_n){d^3q_1d^3q_2\over q_1^{\,0}
q_2^{\,0}}.\label{eeight}\end{eqnarray} 
We may now use this result (\ref{eeight}) in (\ref{efour}) with the
existing DGLAP evolved structure functions~\cite{strfn} without
double counting any effects whatsoever. Moreover, the result 
(\ref{eeight}) is finite and has a smooth limit as 
the initial state light quark
masses $m_q\rightarrow 0$.\par
A related point is that, from the formulas for $\tS_{QCD}$ (eq.(5) in
the first paper in
Ref.~\cite{qcdexp}) and for $\tS_{QCD}|_{DGLAP}$
it is clear that the terms we are exponentiating in the factor
$exp\{SUM_{IR}(QCD)\}$ in (\ref{eone}) are indeed the usual ones
that all published calculations have treated to a finite order in
$\alpha_s$ when they combine soft real radiation cross sections with
virtual gluon radiation corrected ones to cancel the respective infrared
singularities at {\it $k_0\rightarrow 0$} to the respective order in
$\alpha_s$; here, we treat this cancelation to all orders in $\alpha_s$.
Referring to Refs.~\cite{rdfield,eletal,elsex,nas,been}, 
we see that the value of
$\alpha_s$ that should be used in $SUM_{IR}(QCD)$ is $\alpha_s(\mu)$
where $\mu$ is the respective hard renormalization scale$\sim$factorization
scale in the hard process- otherwise, the real gluon emission $k_o\rightarrow 0$ singularity will not cancel the respective 
virtual $k_o\rightarrow 0$ singularity: for illustration, we
recall the result in Ref.~\cite{rdfield} for the parton level
Drell-Yan~\cite{drelly} process $q+\bar q\rightarrow \mu^+\mu^-+G$ for
a muon pair mass $M$ with the IR gluon regulator mass $m_G\rightarrow 0$
, eq.(5.3.17) in Ref.~\cite{rdfield} (here, $\hat\sigma_0$ is the respective
Born cross section),
\begin{equation}
\hat\sigma_{DY}(real)={2\alpha_s(M)\over 3\pi}\hat\sigma_0\{\ln^2(m_G^2/M^2)
+3\ln(m_G^2/M^2)+\pi^2\}
\label{dy1}
\end{equation}
and its corresponding virtual correction, eq.(5.3.18) in Ref.~\cite{rdfield},
\begin{equation}
\hat\sigma_{DY}(virtual)={2\alpha_s(M)\over 3\pi}\hat\sigma_0\{-\ln^2(m_G^2/M^2)
-3\ln(m_G^2/M^2)-{7\over 2}-{2\pi^2\over 3}+\pi^2\},
\label{dy2}
\end{equation}
the two of which sum to the total result, eq.(5.3.19) in Ref.~\cite{rdfield},
\begin{equation}
\hat\sigma_{DY}(real)+\hat\sigma_{DY}(virtual)={2\alpha_s(M)\over 3\pi}\hat\sigma_0
\left\{{4\pi^2\over 3}-{7\over 2}\right\},
\label{dy3}
\end{equation}
which is independent of the infrared $(k_0\rightarrow 0)$ gluon regulator
mass $m_G$ -- the $k_0\rightarrow 0$ gluon infrared singularities regulated
by $m_G$ in (\ref{dy1}) and (\ref{dy2}) are cancelled in (\ref{dy3})
{\it at the hard scattering coupling $\alpha_s(M)$}. We refer the reader to
Refs.~\cite{eletal,elsex}, to the 
second paper in Refs.~\cite{nas}, and to
the second paper in Ref.~\cite{been}
for further explicit examples of this choice of $\alpha_s$ in the 
cancelation of IR singularities at $k_o\rightarrow 0$ in finite
order perturbative QCD calculations. More recently, the 
reader may also see the analysis in
Ref.~\cite{cat} for further illustrations of this point.
Finally, we recall the fundamental analysis of Altarelli and Parisi
in Ref.~\cite{DGLAP} in which the regularized DGLAP kernel, $P_{qq}(z)$,
generates the differential change in the total $qq$ splitting 
function given by
\begin{equation}
d\Gamma_{qq}(z,t) = {\alpha_s(Q)\over 2\pi}P_{qq}(z)dt
\label{gl1}
\end{equation}
and 
\begin{equation}
P_{qq}(z) = C_F{\Big\lgroup}{1+z^2\over (1-z)_+}+{3\over 2}\delta(1-z)
{\Big\rgroup}
\label{gl2}
\end{equation}
where the $+$function has the standard definition
\begin{equation}
\int_0^1dz{f(z)\over (1-z)_+}\equiv \int_0^1dz{f(z)-f(1)\over (1-z)}
\end{equation}
for any appropriate test function $f(z)$, $C_F = 4/3$ is the respective
quark color Casimir invariant and $t = \ln Q^2/m_q^2 -1$ for example.
$Q$ is the respective hard scale.
We focus here on the $\delta(1-z)$
part of $d\Gamma_{qq}(z,t)$ which realizes the cancellation of the
$k_0\rightarrow 0$ real and virtual gluon singularities in  $d\Gamma_{qq}(z,t)$
and which enters into it with the squared coupling constant of
$\alpha_s(Q)$, the coupling constant of the respective
hard scale $Q$, in complete analogy with the value of $\alpha_s$ which
we use in $SUM_{IR}(QCD)$. The correctness of the value of $\alpha_s$
multiplying the delta function in $d\Gamma_{qq}(z,t)$ is proven by
agreement between the predictions of the DGLAP equation and the
Wilson expansion for the deep inelastic structure functions (see Ref.~
\cite{DGLAP,wil,gwp}, for example)
and by the agreement of these predictions with the available data~
\cite{dis}.
Thus, the value of $\alpha_s$ 
which we use in $SUM_{IR}(QCD)$ is well-founded in the published
literature.\par
Up to this point in our discussion, we have focussed on how one
uses the existing DGLAP evolved structure functions in conjunction
with the renormalization group improved YFS theory to compute
the rigorous hadron level cross section. It is also possible to
go one step further, to use the explicit application of
the YFS theory to the one-loop and single bremsstrahlung correction
illustrated in Ref.~\cite{prd1991} to exponentiate the
IR singularities in the DGLAP equation itself. This calculation,
which together with its application
will be reported elswehere~\cite{jwelse}, yields an
entirely new approach to DGLAP evolution in which the
splitting functions themselves are actually YFS exponentiated.\par

A final important issue that can be addressed is the relationship
between our YFS exponentiated result (\ref{eone}) and the soft gluon
resummation theory of Sterman~\cite{ref13} and of 
Catani and Trentadue~\cite{ref14}. We now turn to this relationship.
We note that this latter resummation theory has recently been used in
Refs.~\cite{ref3a,ref15,ref16} at the respective leading log level 
to discuss the effects of soft gluons
in the $p+\bar p \rightarrow t+\bar t +X$ cross section normalization.  
\par
Specifically, referring to the notation of 
Refs.~\cite{ref13,ref14,ref3a,ref15,ref16} we can identify the resummed
soft gluon cross section of the theory of Refs.~\cite{ref13,ref14}
with the cross section in (\ref{eone}) via
\[
 \exp[{\rm SUM_{IR}(QCD)}]\sum_{n=0}^\infty\int\prod_{j=1}^n{d^3
k_j\over k_j}\int{d^4y\over(2\pi)^4}e^{iy\cdot(p_1+p_2-q_1-q_2-\sum k_j)+
D_\rQCD}\] \begin{equation}*\bar\beta_n(k_1,\ldots,k_n){d^3q_1\over q_1^{\,0}}{d^3q_2\over
q_2^{\,0}}/d{\cal \Phi} \quad 
\operatornamewithlimits{\Longrightarrow}_{1-z\rightarrow 0} \quad
 e^{E(1-z,\alpha_s)}d\hat\sigma_B/d{\cal \Phi},
\label{eqa1}
\end{equation}
where we note the limit $1-z\rightarrow 0$ is the limit 
studied by Sterman, Catani and Trentadue in which the hard scattering
invariant mass squared fraction z, defined by $s'/s$,
approaches its maximum value 1,
and where $E(1-z,\alpha_s)$ is the resummation exponent estimated
in Ref.~\cite{ref13} at the leading-log level and in Ref.~\cite{ref14}
at the next-leading log level and {\em used} in Refs.~\cite{ref3a,ref15,ref16}
at the leading log level for the normalization of $\sigma(t\bar t)$ at
FNAL, $d{\cal \Phi}$ is the respective Lorentz invariant phase space
considered in Refs.~\cite{ref13,ref14}
and $d\hat\sigma_B$ is the respective Born level cross section. For example,
the well-known double integral over $\alpha_s(\bar k^2_{\perp})$ in E is easily
recovered from the lefthand side of (\ref{eqa1}) by approximating
\begin{eqnarray}
D_{QCD}(y)&\approx& D_{QCD}(0)=\int {d^3k\over k}\tilde S_{QCD}(k)(1-\theta(K_{max}-k))\nonumber \\
   & =& \int {dz\over 1-z}\int {dk^2_\perp \over k^2_\perp}
{\alpha_s(\bar k^2_\perp)\over \pi}C_F(1-\theta((1-z)_{max}-(1-z)))\{1+\cdots\},
\label{eqa2}
\end{eqnarray}
so that for the leading log initial state part of $SUM_{IR}(QCD)+D_{QCD}(0)$
we recover exactly the double integral over $\alpha_s(\bar k^2_\perp)=\alpha_s((1-z) k^2_\perp)\simeq \alpha_s((1-z)Q^2)$
proposed for E in Refs.~\cite{ref13,ref14} (the $\cdots$ in (\ref{eqa2})
stand for the remainder of $D_{QCD}(0)$~\cite{qcdexp}). Here, in our
infinite momentum frame (light-cone) kinematics
, we identified $(1-z)_{max}=K_{max}/p_1^0$ and $Q^2=s$.
We note that, as the
authors in Refs.~\cite{ref13,ref14} have pointed-out,
in allowing $\alpha_s$ to run in (\ref{eqa2}) one resums a certain class
of leading logs which in our exact series in $\alpha_s(Q^2)$ in (\ref{eone}) are
generated by the infinite sum on $n$ in (\ref{eone}). We note
further that the remainder of $E$ corresponds to the next leading
log terms generated by the infinite sum on $n$ in (\ref{eone}) over the
$\bar\beta_n$ when one makes the approximation
\begin{eqnarray}
\sum_{n=0}^\infty {1\over n!}\int\prod_j^n {d^3k_j\over k_j}\delta(p_1+p_2-q_1-q_2-\sum k_j)
\bar\beta'_n(k_1,\cdots,k_n){d^3q_1\over q_1^{\,0}}{d^3q_2\over
q_2^{\,0}}/d\hat\sigma_B \equiv  {\cal S}(\bar\beta') \nonumber\\
\qquad \qquad \approx \sum_{n=0}^\infty {1\over n!}{(E^{nll})}^n \nonumber\\
\label{eqa3}
\end{eqnarray}
where we define $E^{nll}$ to be the next leading log part of $E$ and to be
consistent we have defined $\bar\beta'_n$ to be that part of the
$\bar\beta_n$ in our exact result (\ref{eone}) whose leading log
content is not already resummed by integral over
the running $\alpha_s$ in (\ref{eqa2}).
We see that in our approach, the results of Refs.~\cite{ref13,ref14} for
$E^{nll}$ correspond to the approximation
\begin{equation}
(E^{nll})^n = \int\prod_j^n {d^3k_j\over k_j}\delta(p_1+p_2-q_1-q_2-\sum k_j)
\bar\beta'_n(k_1,\cdots,k_n){d^3q_1\over q_1^{\,0}}{d^3q_2\over
q_2^{\,0}}/d\hat\sigma_B.
\label{eqa3b}
\end{equation}
We conclude that, at the parton level, the formulas in 
Refs.~\cite{ref13,ref14} are entirely contained in our 
formula (\ref{eone}) and that they represent
an approximation to our result (\ref{eone}) in the limit 
that $1-z\rightarrow 0$.\par
One of the important consequences of the identification of the
Sterman-Catani-Trentadue exponent $E$ in (\ref{eqa1}) is that we can
address the issue of the presence or absence of renormalon~\cite{renorm} 
behavior
associated with the leading log and next-leading log truncation of
our exact result (\ref{eone}) as it is reperesented by $E$.
Specifically, as the authors in Refs.~\cite{ref13,ref14,ref3a,ref15,ref16}
have noted, as $1-z\rightarrow 0$ in (\ref{eqa2}), the argument of
the running $\alpha_s$ would appear to approach the Landau pole, creating an
ambiguity in the non-leading terms in the exponent, for example, of the
famous renormalon type~\cite{renorm}{\em if one expands $\alpha_s$ in terms
of this one-loop result}-- we stress that the two and three loop formulas
are known and they do not seem to support this pole in $\alpha_s$
~\cite{brodsky}. On the other hand, we have
argued in Ref.~\cite{qcdexp}, by the uncertainty principle, that the
truly long wavelength gluons , with wavelengths much larger than
$1/\Lambda_{QCD}$, should decouple from the hard process under 
discussion here. Does this decoupling really happen or not?
We look at the exact initial state formula for the sum
\begin{align}
SUM_{IR}(QCD)|_{initial~state}&=\{2\alpha_s ReB_{QCD}+2\alpha_s\tilde B_{QCD}(\Kmax)\}|_{initial~state} \notag \\
             &= -{C_F\over 4\pi^2}\int{d^3k\over k^0}\alpha_s(\bar k^2_\perp)
\lgroup {p_1\over p_1k} - {p_2\over p_2k} \rgroup^2\theta(\Kmax -k)\notag\\
&+ \Re{iC_F\over 4\pi^2}\int{d^4k\over \pi}\lgroup{1\over k^2-m^2_G+i\epsilon}\rgroup \Big\lgroup {2p_1+k\over k^2+2p_1k+i\epsilon}\notag\\&\qquad+
{2p_1+k\over k^2+2p_1k+i\epsilon}\Big\rgroup^2\alpha_s(\bar k^2_\perp)\notag\\
\label{eqa10}
\end{align} 
where we have restored the running ansatz of Refs.~\cite{ref13,ref14}
in their estimate of $E$. We stress that for consistency the same
definition of $\alpha_s$ has to be used for both soft real and soft virtual
gluons in order for the infrared singularities to cancel. Carrying out the
virtual $k^0$ integral by standard contour methods in the upper 
complex $k^0$ plane, we get, for the residue of the gluon propagator
$\equiv Res(gluon~propagator)$,
the contribution 
\begin{equation}
2\alpha_s ReB_{QCD}|_{initial~state, Res(gluon~propagator)}=
{C_F\over 4\pi^2}\int{d^3k\over k^0}\alpha_s(\bar k^2_\perp)
\lgroup {p_1\over p_1k} - {p_2\over p_2k} \rgroup^2
\label{eqa11}
\end{equation}
so that, when we combine this result with the real emission
term in (\ref{eqa10}), we get finally the representation
\begin{align}
SUM_{IR}(QCD)|_{initial~state}&={C_F\over 4\pi^2}\int_{k^0\ge K_{max}}
{d^3k\over k^0}
\alpha_s(\bar k^2_\perp)\lgroup {p_1\over p_1k} - {p_2\over p_2k} \rgroup^2
\notag\\
&-{C_F\over 2\pi}\Re\int{d^3k\over \pi}\sum_{Res(fermion~propagators)}
\lgroup{1\over k^2-m^2_G+i\epsilon}\rgroup\notag\\ & \qquad \times 
\Big\lgroup {2p_1+k\over k^2+2p_1k+i\epsilon} +
{2p_1+k\over k^2+2p_1k+i\epsilon}\Big\rgroup^2\alpha_s(\bar k^{\prime 2}_\perp),
\notag\\
\label{eqa12}
\end{align}   
where we have introduced the scale $\bar k^{\prime 2}_\perp$ for $\alpha_s$ 
in the second
term in this last equation to take into account that, as this term is
not infrared divergent and is dominated by gluon momenta ${\cal O}(\sqrt s/2)$,
$\bar {k^\prime}_\perp$ should also be of this order for consistency. 
In the second term in (\ref{eqa12}), only the fermion propagator residues
, $Res(a), a= fermion~propagators$, enter into the respective sum over
residues, as we indicate explicitly.
We see in the first
term in (\ref{eqa12}) that only gluons with energies exceeding the dummy
parameter $K_{max}$ (our result (\ref{eone}) 
is independent of $K_{max}$ and $1-z_{max}$) are
actually involved in $SUM_{IR}(QCD)$ so that, if we use the resummation
ansatz of Refs.~\cite{ref13,ref14} we do not encounter the regime
$(1-z)s=\Lambda_{QCD}^2$ in our result for the LL part of $E$ as we
may take $(1-z_{max})\gg {\Lambda_{QCD}^2\over s}$ without loss of content
or predictive power. This is consistent with the earlier observation that
(\ref{eone}), which is a series in $\alpha_s(Q^2)$, is consistent with the
uncertainty principle and hence 
in fact is insensitive
to the behavior of $\alpha_s$ at scales ${\cal O}(\Lambda_{QCD})$
when $Q^2\gg \Lambda_{QCD}^2$.
\par
In summary, we have shown how one rigorously synthesizes the powerful
results of DGLAP evolution and renormalization group improved
YFS exponentiation in QCD without
double counting. Explicit Monte Carlo event generator data based on
our prescription will appear elsewhere~\cite{elswh}.\par

\vspace{0.5cm}
\noindent{\bf Acknowledgments}
We would thank Profs. G. Veneziano and
G. Altarelli for the support and kind hospitality of the CERN
Theory Division, where a part of this work was performed. One of
us (B.F.L. W.) thanks Prof. C. Prescott of SLAC for the kind hospitality
of SLAC Group A while this work was completed.
 


\end{document}